\documentclass{amsart}
\usepackage{graphicx,amssymb}
\newtheorem{thm}{Theorem}[section]
\newtheorem{cor}[thm]{Corollary}

\theoremstyle{definition}

\theoremstyle{remark}

\newcommand{\mbold}[1]{\mbox{\boldmath${#1}$}}
\def\beq{\begin{eqnarray}}
\def\eeq{\end{eqnarray}}
\def\d{\mathrm{d}}
\def\Wm{\mathcal{W}_1}
\def\Wc{\mathcal{W}_2}
\begin{document}

\title[The late-time behaviour of vortic Bianchi type VIII Universes]{The late-time behaviour of vortic \\ Bianchi type VIII Universes}%
\author{Sigbj{\o}rn Hervik and  Woei Chet Lim}
\address{Department of Mathematics \& Statistics, Dalhousie University,
Halifax, Nova Scotia,
Canada B3H 3J5  }%
\email{herviks@mathstat.dal.ca, wclim@mathstat.dal.ca}%

\date{\today}
\begin{abstract} We use the dynamical systems approach to investigate the Bianchi type VIII models with a tilted $\gamma$-law
perfect fluid.  We introduce expansion-normalised variables and investigate the late-time asymptotic behaviour 
of the models and
determine the late-time asymptotic states. For the Bianchi type VIII models the state space is unbounded and consequently, for all non-inflationary perfect fluids, one of the curvature variables grows without bound. Moreover, we show that for fluids stiffer than dust ($1<\gamma<2$), the fluid will in general tend towards a state of extreme tilt. For dust ($\gamma=1$), or for fluids less stiff than dust ($0<\gamma< 1$), we show that the fluid will in the future be asymptotically non-tilted. Furthermore, we show that for all $\gamma\geq 1$ the universe evolves  towards a vacuum state but does so rather slowly, $\rho/H^2\propto 1/\ln t$.  
\end{abstract} 
\maketitle 

\section{Introduction}
Cosmology in the recent years has proven to be an important arena to perform tests of the general theory of relativity. Today, theoretical cosmology gives insights into the possible physical and mathematical properties of the universe while observations help to constrain the various theoretical possibilities. The purpose of this paper is to fill  one of the major gaps in the theoretical understanding of the behaviour of spatially homogeneous (SH) cosmologies \cite{EM,BS,DS1,DS2}. The aim is to give a description of the evolution of a general spatially homogeneous model with a perfect fluid, in particular, the  SH Bianchi type VIII models with a $\gamma$-law perfect fluid. Until now, only Bianchi type VIII models where the fluid flow is orthogonal to the surfaces of homogeneity have been studied. More specifically, vacuum Bianchi type VIII models were studied in \cite{BG,Ringstrom1,Ringstrom2} and a detailed derivation of the asymptotic expansions of Bianchi type VIII vacuum metrics was given in \cite{Ringstrom3}. The future asymptotic behaviour with a non-tilted perfect fluid was also studied in \cite{VIII}. Here, we will allow for the fluid flow to be tilted; i.e., where the fluid flow is not orthogonal to the surfaces of homogeneity \cite{KingEllis}.

Several SH models with tilt have been studied before: type II \cite{HBWII}, IV \cite{CH,HHC}, V \cite{Shikin,Collins,CollinsEllis,HWV,Harnett}, VI$_0$ \cite{hervik,coleyhervik}, VII$_0$ \cite{HHLC}, and VII$_h$ \cite{CH,HHC} (see also \cite{CH,BHtilted} for a subclass of the type VI$_h$ models). 
 Allowing for a tilted fluid, results in several interesting new phenomena, such as  future limiting curves and attracting tori \cite{HHC}. Most of the above-mentioned works have been utilizing the theory of dynamical systems in their investigations \cite{DS1,DS2,BN}; we will do the same by generalising the formalism from the solvable case \cite{CH} to the semisimple Bianchi type VIII models.

The Bianchi type VIII models are particularly interesting models and are the most general ever-expanding cosmological model of Bianchi type. The state space of the tilted $\gamma$-law type VIII models is of dimension 8 (compared to 5 in the non-tilted case and 4 in the vacuum case); hence, by studying the type VIII models we can gain insight into the behaviour of a `general' Bianchi model. 

Under many physical circumstances there are certain  self-similar solutions that act as attractors for more general solutions of the model. This `self-similarity hypothesis' \cite{CC1,CC2} is known to be valid for many SH models, however, there are a few notable exceptions. It was pointed out in \cite{VII0} that for the non-tilted perfect fluid Bianchi type VII$_0$ models one of the curvature variables grows without bound and the models are consequently not asymptotically self-similar. The same is also true for the non-tilted perfect fluid type VIII models \cite{VIII}. By allowing for a tilted perfect fluid we will see that this property does not change. Neither the tilted type VII$_0$ models (which was shown in \cite{HHLC}) nor the tilted type VIII models are asymptotically self-similar. 

The Bianchi type VIII Lie algebra corresponds to the Lie algebra of the matrix group $SL(2,\mathbb{R})$ with a connected covering group usually denoted by $\widetilde{SL(2,\mathbb{R})}$. This group is one of the eight so-called \emph{Thurston geometries} (see, e.g. \cite{thurston97}). The Thurston geometries play an important role in the famous geometrization conjecture for 3-manifolds by Thurston \cite{thurston}\footnote{See, e.g. \cite{Morgan} for a review of the geometrization conjecture and a discussion of a recent attempt of proving it.}. The type VIII model is usually considered to be the connected and simply connected group $\widetilde{SL(2,\mathbb{R})}$ (we will do the same here); however, it is worth pointing out that the group  $\widetilde{SL(2,\mathbb{R})}$ allows for a discrete subgroup $\Gamma$ such that the quotient $\widetilde{SL(2,\mathbb{R})}/\Gamma$ is compact. In this regard Barrow and Kodama \cite{BK1,BK2} pointed out that as the complexity of the quotient $\widetilde{SL(2,\mathbb{R})}/\Gamma$ increases, the number of parameters describing the corresponding Bianchi model grows without bound. 

A set of left-invariant one-forms on $\widetilde{SL(2,\mathbb{R})}$ can, for example, be given by:
\beq
{\mbold\omega}^1&=&a\left(\d x-\frac{\d z}{y}\right), \nonumber \\
{\mbold\omega}^2&=&\frac by\left(\cos x\ \d y+\sin x\ \d z\right), \nonumber \\
{\mbold\omega}^3&=&\frac cy\left(-\sin x\ \d y+\cos x\ \d z\right).
\label{VIII1forms}\eeq
These one-forms fulfil the type VIII Lie algebra relations:
\beq
\d {\mbold\omega}^i=-\frac 12C^i_{jk}{\mbold\omega}^j\wedge{\mbold\omega}^k, \quad C^i_{jk}=\varepsilon_{jkl}n^{li},
\eeq
where $n^{ij}$ is a symmetric matrix with eigenvalues $\lambda_1<0<\lambda_2,\lambda_3$. For the choice (\ref{VIII1forms}), 
\beq
(n^{ij})=\mathrm{diag}\left(-\frac{a}{bc},\frac{b}{ac},\frac{c}{ab}\right). 
\eeq
Using the one-forms (\ref{VIII1forms}) we can introduce a Bianchi type VIII cosmology by
\beq
\d s^2=-\d t^2+\delta_{ab}\widetilde{\mbold\omega}^a\widetilde{\mbold\omega}^b, \quad \widetilde{\mbold\omega}^a={\sf e}^a_{~i}(t){\mbold\omega}^i.
\eeq
Here, $t$ is the proper time of an observer whose world-line is a geodesic orthogonal to the type VIII surfaces of homogeneity (also called the cosmological time). The hypersurface orthogonal vector is ${\bf n}=\partial/\partial t$.  

The energy-momentum tensor of the perfect fluid is 
\beq
T_{\mu\nu}=(p+\rho)u_{\mu}u_{\nu}+pg_{\mu\nu},
\eeq
where $\rho$, $p$ and $u^{\mu}$ are the energy density, pressure and four-velocity of the fluid, respectively. The equation of state will be taken to be 
\beq
p=(\gamma-1)\rho,
\eeq 
where $\gamma$ is a constant. This choice includes the important cases of dust ($\gamma=1$) and radiation ($\gamma=4/3$). 
In this paper we will study models where the four-velocity $u^{\mu}$ of the perfect fluid is not parallel with the normal vector $n^{\mu}$. Since these models are future geodesically complete \cite{Rendall}, we will choose the fundamental observers to follow the geodesic congruences defined by the vector field $n^{\mu}$. This choice avoids the possible singular behaviour that may occur for observers following the fluid flow lines \cite{CollinsEllis,CHL}.

\section{Equations of motion for the tilted Bianchi type VIII models}
Using the orthonormal frame formalism we introduce expansion-normalised variables and write the Einstein equations as an autonomous system of differential equations with constraints \cite{DS1,DS2}. We will follow the formalism of \cite{CH} carefully with only minor differences.
We introduce a dimensionless time variable, $\tau$, defined by: 
\[ \frac{\d t}{\d \tau}=\frac 1H,\]
where $H$ is the Hubble scalar defined by $H=(1/3)n^{a}_{~;a}$. Moreover, the shear tensor, $\sigma_{ab}$, is the trace-free part of $n_{a;b}$: $\sigma_{ab}=n_{a;b}-H(g_{ab}+n_an_b)$.  
We parameterise the expansion-normalised $N_{ab}$ and shear $\Sigma_{ab}$  as:
\beq 
(N_{ab})&=&\sqrt{3}\begin{bmatrix}
N_{1} & 0 & 0 \\
0 & \bar{N}+N_- & N_{23} \\
0 & N_{23} & \bar{N}-N_-
\end{bmatrix}\nonumber \\
(\Sigma_{ab})&=&\begin{bmatrix}
-2\Sigma_+ & \sqrt{3}\Sigma_{12} & \sqrt{3}\Sigma_{13} \\
 \sqrt{3}\Sigma_{12}& \Sigma_++\sqrt{3}\Sigma_- & \sqrt{3}\Sigma_{23} \\
\sqrt{3}\Sigma_{13} & \sqrt{3}\Sigma_{23} & \Sigma_+-\sqrt{3}\Sigma_-
\end{bmatrix}.
\eeq
Following the approach in \cite{CH} we then introduce the complex variables: 
\beq 
&&{\bf N}_{\times}=N_-+iN_{23}, \quad {\mbold\Sigma}_{\times}=\Sigma_-+i\Sigma_{23}, \nonumber \\
&&{\mbold\Sigma}_1=\Sigma_{12}+i\Sigma_{13}, \quad {\bf v}=v_2+iv_3.
\eeq
We still have one free gauge function left which represents a complex rotation as follows:
\beq
\phi:~\left({\bf N}_{\times},\mbold{\Sigma}_{\times},{\mbold\Sigma}_1,{\bf v}\right)~\mapsto ~ \left(e^{2i\phi}{\bf N}_{\times},e^{2i\phi}\mbold{\Sigma}_{\times},e^{i\phi}{\mbold\Sigma}_1,e^{i\phi}{\bf v}\right).
\eeq
Let $R_a$ be the (expansion-normalised) local angular velocity of a Fermi-propagated axis with respect to the triad ${\bf e}_a$. We will use the gauge function and replace $R_1$  with $\phi'$ in the equations of motion. Various choices of $\phi$ correspond to different choices of gauge and we will choose $\phi'=0$ for most of this paper. This leaves us with a remaining constant gauge rotation (see \cite{CH} for a discussion of the choices of gauge). However, other choices may be convenient for certain purposes; for example, for the numerical runs we have chosen $\mathrm{Re}({\mbold\Sigma}_1)=0$.  

We introduce $R_3-iR_2=\sqrt{3}{\bf R}$ where
\beq~
{\bf R}=a{\mbold\Sigma}_1+b{\mbold\Sigma}_1^*,
\quad a=\frac{\bar{N}^2-N_1^2-\left|{\bf N}_{\times}\right|^2}{(\bar{N}-N_1)^2-\left|{\bf N}_{\times}\right|^2}, \quad  b=\frac{-2N_1{\bf N}_{\times}}{(\bar{N}-N_1)^2-\left|{\bf N}_{\times}\right|^2}.
\eeq
For the tilted type VIII models both $a$ and $b$ are bounded, $|a|\leq 1$, $|b|\leq 1$, and are thus well-defined.  

The equations of motion are:
\beq \Sigma_+'&=&
(q-2)\Sigma_++{3}\mathrm{Re}\left({\bf R}^*{\mbold\Sigma}_1\right)-N_1\bar{N}-2|{\bf N}_{\times}|^2
+\frac{\gamma\Omega}{2G_+}\left(-2v_1^2+|{\bf v}|^2\right) \label{eq:Sigma+eq}\\
{\mbold\Sigma}_{\times}'&=&
(q-2+2i\phi'){\mbold\Sigma}_{\times}+\sqrt{3}{\mbold\Sigma}_{1}{\bf R}
-\sqrt{3}{\bf N}_{\times}(2\bar{N}-N_1)
+\frac{\sqrt{3}\gamma\Omega}{2G_+}{\bf v}^2
\\
{\mbold\Sigma}'_{1}&=& \left(q-2+i\phi'\right){\mbold
\Sigma}_{1} -3\Sigma_+{\bf R}-\sqrt{3}{\mbold\Sigma}_{\times}{\bf R}^*
+\frac{\sqrt{3}\gamma\Omega v_1}{G_+}{\bf v}
\\
 {\bf N}_{\times}'&=& \left(q+2\Sigma_++2i\phi'\right){\bf N}_{\times}+2\sqrt{3}{\mbold\Sigma}_{\times}\bar{N}\\
\bar{N}'&=&
\left(q+2\Sigma_+\right)\bar{N}+2\sqrt{3}\mathrm{Re}\left({\mbold\Sigma}_{\times}^*{\bf
N}_{\times}\right)\\
N_1' &=& (q-4\Sigma_+)N_1
 \label{eq:Aeq}\eeq
The equations for the fluid are
\beq
\quad \Omega'&=& \frac{\Omega}{G_+}\Big\{2q-(3\gamma-2)
 +\left[2q(\gamma-1)-(2-\gamma)-\gamma\mathcal{S}\right]V^2\Big\}\label{eq:Omega}
 \quad \\
 v_1' &=& \left(T+2\Sigma_+\right)v_1-\sqrt{3}\mathrm{Re}\left[({\bf R}+{\mbold\Sigma}_{1}){\bf v}^*\right] +\sqrt{3}\mathrm{Im}({\bf N}_{\times}^*{\bf v}^2)\\
 {\bf v}'&=& \left[T-\Sigma_+-i\sqrt{3}(\bar{N}-N_1)v_1+i\phi'\right]{\bf v}-\sqrt{3}\left({\mbold\Sigma}_{\times}+i{\bf N}_{\times}v_1\right){\bf v}^*\nonumber \\ && -\sqrt{3}({\mbold\Sigma_1}-{\bf R})v_1 \\
  V'&=&
\frac{V(1-V^2)}{1-(\gamma-1)V^2}\left[(3\gamma-4)-\mathcal{S}\right] \label{eq:Veq}
\eeq where
\beq q&=& 2\Sigma^2+\frac
12\frac{(3\gamma-2)+(2-\gamma)V^2}{1+(\gamma-1)V^2}\Omega\nonumber \\
\Sigma^2 &=& \Sigma_+^2+|{\mbold{\Sigma}}_{\times}|^2+|{\mbold\Sigma}_{1}|^2\nonumber \\
\mathcal{S} &=& \Sigma_{ab}c^ac^b, \quad c^ac_{a}=1, \quad v^a=Vc^a,\quad \nonumber \\
 V^2 &=& v_1^2+|{\bf v}|^2,\quad  \nonumber \\
G_+ &=& 1+(\gamma-1)V^2, \nonumber \\
 T&=& \frac{(3\gamma-4)(1-V^2)+(2-\gamma)V^2\mathcal{S}}{1-(\gamma-1)V^2}.
\label{eq:defs}
\eeq
These variables are subject to the constraints
\beq
1&=& \Sigma^2+\frac 14N_1^2-N_1\bar{N}+|{\bf N}_{\times}|^2+\Omega \label{const:H}\\
0 &=& 2\mathrm{Im}({\mbold\Sigma}_{\times}^*{\bf N}_{\times})+\frac{\gamma\Omega v_1}{G_+} \label{const:v1}\\
0 &=&
i\mbold{\Sigma}_{1}(\bar{N}-N_1)+i{\mbold{\Sigma}}^*_{1}{\bf
N}_{\times}+\frac{\gamma\Omega {\bf v}}{G_+} \label{const:v2}
\eeq 
The
parameter $\gamma$ will be assumed to be in the interval $\gamma\in (0,2)$ and we will henceforth choose the `F-gauge', $\phi'=0$.  The generalised
Friedmann equation, eq.(\ref{const:H}), yields an expression which effectively determines the
energy density $\Omega$.  The state vector will therefore be considered to be ${\sf
X}=[\Sigma_+,{\mbold\Sigma}_{\times},{\mbold\Sigma}_1,{\bf N}_{\times},\bar{N},N_1,v_1,{\bf v}]$
modulo the constraint equations (\ref{const:v1}) and (\ref{const:v2}).  

The dynamical system is invariant under the following discrete symmetries: 
\begin{align}
\phi_1: & ~ [\Sigma_+,{\mbold\Sigma}_{\times},{\mbold\Sigma}_1,{\bf N}_{\times},\bar{N},N_1,v_1,{\bf v}]\mapsto [\Sigma_+,{\mbold\Sigma}_{\times},{\mbold\Sigma}_1,-{\bf N}_{\times},-\bar{N},-N_1,-v_1,-{\bf v}]\nonumber \\
\phi_2: & ~[\Sigma_+,{\mbold\Sigma}_{\times},{\mbold\Sigma}_1,{\bf N}_{\times},\bar{N},N_1,v_1,{\bf v}]\mapsto [\Sigma_+,{\sf u}^2{\mbold\Sigma}_{\times}^*,-{\sf u}{\mbold\Sigma}_1^*,{\sf u}^2{\bf N}_{\times}^*,\bar{N},N_1,-v_1,{\sf u}{\bf v}^*], ~~{\sf u}\in S^1\nonumber \\ 
\phi_3: &~[\Sigma_+,{\mbold\Sigma}_{\times},{\mbold\Sigma}_1,{\bf N}_{\times},\bar{N},N_1,v_1,{\bf v}]\mapsto [\Sigma_+,{\mbold\Sigma}_{\times},-{\mbold\Sigma}_1,{\bf N}_{\times},\bar{N},N_1,v_1,-{\bf v}]\nonumber.
\end{align}
These discrete symmetries imply that without loss of generality we can, for the type VIII models, restrict the 
variables $\bar{N}> 0$, $N_1<0$ and $v_1\geq 0$. \footnote{There is a subtlety regarding this 
choice since, in general, $v_1=0$  is not an invariant subspace. The state space can be 
considered an orbifold with a mirror symmetry at $v_1=0$; in particular, this means that 
any equilibrium point in the region $v_1>0$ has an  analogous equilibrium point in the 
region $v_1<0$.}  We also note that the free parameter ${\sf u}$ in $\phi_2$ is, in fact, 
the remaining gauge transformation. 

\subsection{Invariant Subspaces}In this analysis we will be concerned with the following invariant sets:
\begin{enumerate}
\item{} $T(VIII)$: The general tilted type VIII model with $N_1<0$, $\bar{N}^2-\left|{\bf N}_{\times}\right|^2>0$.
\item{} $N(VIII)$: A tilted type VIII model defined by $N_1<0$, $\bar{N}^2-\left|{\bf N}_{\times}\right|^2>0$, $N_1v_1^2+\bar{N}|{\bf v}|^2+\mathrm{Re}({\bf N}_{\times}^*{\bf v}^2)=0$
\item{} $T_1(VIII)$: A one-tilted type VIII model defined by $N_1<0$, $\bar{N}^2-\left|{\bf N}_{\times}\right|^2>0$, ${\mbold\Sigma_1}={\bf v}=0$. 
\item{} $F(VIII)$: The set of fixed points of $\phi_2$ (without loss of generality we 
can set ${\sf u}=1$) defined by $v_1=\mathrm{Im}({\mbold\Sigma}_{\times})=\mathrm{Re}({\mbold\Sigma}_1)=\mathrm{Im}({\bf N}_{\times})=\mathrm{Im}({\bf v})=0$.
\item{} $B(VIII)$: Non-tilted Bianchi type VIII models with $N_1<0$, $\bar{N}^2-\left|{\bf N}_{\times}\right|^2>0$, $v_1={\bf v}=0$.
\item{} $T(VI_0)$: The general tilted type VI$_0$ model with $N_1<0$, $\bar{N}=\left|{\bf N}_{\times}\right|>0$.
\item{} $T(VII_0)$: The general tilted type VII$_0$ model with $N_1=0$, $\bar{N}^2-\left|{\bf N}_{\times}\right|^2>0$.
\item{} $T(II_{(1)})$: The general tilted type II model given by $N_1<0$, $\bar{N}=\left|{\bf N}_{\times}\right|=0$.
\item{} $T(II_{(2)})$: The general tilted type II model given by $N_1=0$, $\bar{N}=\left|{\bf N}_{\times}\right|>0$.
\item{} $B(I)$: Type I: $\bar{N}={\bf N}_{\times}=N_1=V=0$.
\item{} $\partial T(I)$: ``Tilted'' vacuum type I: $\Omega=\bar{N}={\bf N}_{\times}=N_1=0$.
\end{enumerate}
We note that the closure of the set $T(VIII)$ is given by:
\beq
\overline{T(VIII)}=T(VIII)\cup T(VI_0)\cup T(VII_0)\cup T(II_{(1)})\cup T(II_{(2)})\cup B(I)\cup \partial T(I).\nonumber 
\eeq
Moreover, in $\overline{T(VIII)}$ we have the bounds
\beq
\Sigma_+^2+|{\mbold{\Sigma}}_{\times}|^2+|{\mbold\Sigma}_{1}|^2+\frac 14N_1^2-N_1\bar{N}+\left|{\bf N}_{\times}\right|^2\leq 1, \quad v^2_1+|{\bf v}|^2\leq 1;
\eeq
hence, all variables, except for  $\bar{N}$, are bounded. The curvature variable $\bar{N}$ can grow without bound, provided that  $|N_1\bar{N}|\leq 1$.  

\subsection{Fluid Vorticity}
The various invariant subspaces can also be categorised in terms of
the ($H_{\mathrm{fluid}}$-normalised) fluid vorticity, $W^{\alpha}$. The vorticity of the fluid for the type VIII model is given by:
\beq
W_a=\frac{1}{2B}\left(N_{ab}v^b+\frac 1{1-V^2}N_{bc}v^bv^cv_a\right), \quad W_0=-v^aW_a, 
\eeq
where 
\[ B=\frac{1-\frac 13(1+\mathcal{S})V^2}{G_-\sqrt{1-V^2}}.\]
\begin{enumerate}
\item{} $T(VIII)$: General vortic type VIII where all components $W^{\alpha}$ can be non-zero.
\item{} $N(VIII)$: $W^0=0$. 
\item{} $T_1(VIII)$: $W^2=W^3=0$.
\item{} $F(VIII)$: $W^1=W^3=0$.
\item{} $B(VIII)$: $W^0=W^a=0$, non-vortic.
\end{enumerate}
\subsection{Monotonic functions}
There are three monotonic functions which are useful for our analysis:
\beq
Z_1& \equiv &\alpha\Omega^{1-\Gamma}, \quad \alpha=\frac{(1-V^2)^{\frac 12(2-\gamma)}}{G_+^{1-\Gamma}V^\Gamma}, \quad \Gamma=\frac 67\gamma, \\
Z_1' &=& \left[2(1-\Gamma)q+(2+2\Gamma-3\gamma)+\Gamma\mathcal{S}\right]Z_1.\nonumber 
\eeq
We note that (by the same trick as in \cite{CH})
\beq 
&&2(1-\Gamma)q+(2+2\Gamma-3\gamma)+\Gamma\mathcal{S} \nonumber \\
&& \geq \frac 17(14-15\gamma)\left(2\Sigma^2+\frac 14N_1^2-N_1\bar{N}+|{\bf N}_{\times}|^2\right)+\frac{\gamma\left[18(1-\gamma)+(10-9\gamma)V^2\right]}{7G_+}\Omega.\nonumber
\eeq
Thus $Z_1$ is monotonically increasing in $T(VIII)$ for $\gamma\leq 14/15$. 

\beq
Z_2&\equiv & N_1^2\left(\bar{N}^2-\left|{\bf N}_\times\right|^2\right)^2, \\
Z_2' &=& 6q Z_2.\nonumber 
\eeq 
$Z_2$ is monotonically increasing for $2/3<\gamma<2$.

\beq
Z_3 &\equiv&\frac{\left[N_1v_1^2+\bar{N}|{\bf v}|^2+\mathrm{Re}({\bf N}_{\times}^*{\bf v}^2)\right]^2}{(1-V^2)^{2(2-\gamma)}\beta\Omega},\quad \beta\equiv\frac{(1-V^2)^{\frac12(2-\gamma)}}{G_+} \\
Z_3' &=&3(5\gamma-6) Z_3. \nonumber
\eeq
Thus $Z_3$ is monotonically decreasing ($\gamma <6/5$) or increasing ($6/5<\gamma$) in $T(VIII)\setminus N(VIII)$.   

\section{Qualitative analysis}
Using the above monotonic functions we obtain some results regarding the asymptotic behaviour of tilted Bianchi type VIII universes: 

\begin{thm}
For $2/3\leq \gamma<2$ all Bianchi models  of type VIII have:
 \[ \lim_{\tau\rightarrow\infty}\left|\bar{N}\right|=\infty.\]
\label{thmNbar}\end{thm}
\begin{proof}
The proof uses $Z_2$ and is similar to the non-tilted type VIII analysis \cite{VIII}. 
\end{proof}
\begin{cor}
For $2/3\leq \gamma<2$ all Bianchi models  of type VIII have:
 \[ \lim_{\tau\rightarrow\infty}({\mbold\Sigma}_1,N_1)=(0,0).\]
\end{cor}
\begin{proof}
From the constraint equations we have that both ${\mbold\Sigma}_1\bar{N}$ and $N_1\bar{N}$ are bounded. Hence, since $\bar{N}\rightarrow \infty$, the corollary follows. 
\end{proof}
Another interesting quantity is $D\equiv N_1v_1^2+\bar{N}|{\bf v}|^2+\mathrm{Re}({\bf N}_{\times}^*{\bf v}^2)$. We note that $D$ can have either sign and $D=0$ defines the invariant subspace $N(VIII)$. The equation of motion for $D$ is particularly simple: 
\[ D'=(q+2T)D.\] 
\begin{thm}
Assume that $2/3\leq \gamma<2$ and that there exists a time $\tau_0$ such that $D(\tau_0)\leq 0$. Then 
\[ \lim_{\tau\rightarrow\infty} {\bf v}=0.\]
\end{thm}
\begin{proof}
Since $D=0$ is an invariant subspace the sign of $D$ cannot change during the evolution of the models. Since $\bar{N}\rightarrow \infty$ while all the other variables are bounded, $D\leq 0$ implies ${\bf v}\rightarrow 0$. 
\end{proof}
\begin{thm}
For $0<\gamma<1$ all tilted Bianchi models (with $\Omega>0$ and $V<1$) of type VIII are asymptotically non-tilted at late times.  
\label{thmV0}\end{thm}
\begin{proof} The proof splits in two cases depending on the value of $\gamma$: \textit{(i)} $0<\gamma<14/15$:
 The theorem follows immediately from the monotonic function $Z_1$. 
\textit{(ii)} $2/3<\gamma<1$: Since $\bar{N}\rightarrow \infty$ the monotonic function $Z_3$ implies ${\bf v}\rightarrow 0$. To show that $v_1\rightarrow 0$ also, we can use the reduced system which we come to later in section \ref{sect:redsys}. For the reduced system we can use the monotonic function $R_2$ (this function is also monotonically increasing in the invariant subspace $T_1(VIII)$) which effectively can be used as a monotonically increasing function for the full system at sufficiently late times. This implies $v_1\rightarrow 0$. 
\end{proof}
\begin{cor}[cosmic no-hair]
For $\Omega>0$, $V<1$, and $0<\gamma<2/3$ we have that
\[ \lim_{\tau\rightarrow\infty}\Omega=1, \quad \lim_{\tau\rightarrow\infty}V=0.\]
\end{cor}
\begin{proof}
See \cite{CH}.
\end{proof}
\subsection{Equilibrium points} 
Using $Z_2$ we note that all equilibrium points have either $Z_2=0$ or $q=0$. The equation $Z_2=0$ defines the Bianchi types VII$_0$, VI$_0$, II and I. The equilibrium points for these Bianchi types are given elsewhere. All of these equilibrium points are unstable into the future for $2/3< \gamma<2$. For $2/3<\gamma<2$, the equation $q=0$ automatically implies $Z_2=0$, and hence, we can conclude: 
\begin{cor}
There are no equilibrium points in $T(VIII)$.
\end{cor} 

\subsection{Late time analysis} 
For the type VIII model it is convenient to solve constraint (\ref{const:v2}) to obtain an expression for ${\mbold\Sigma}_1$:
\beq
{\mbold\Sigma}_1=\frac{i\gamma\Omega}{G_+\left[(\bar{N}-N_1)^2-\left|{\bf N}_{\times}\right|^2\right]}\left[(\bar{N}-N_1){\bf v}+{\bf N}_{\times}{\bf v}^*\right].
\eeq

In the following we will introduce the variables $M$ and $\eta$
\beq
M\equiv \frac{1}{\bar{N}}, \quad \eta\equiv-N_1\bar{N}, \quad 0\leq\eta\leq 1.
\eeq
In light of Theorem \ref{thmNbar}, we have that $M\rightarrow 0$; in particular, 
\beq
{\mbold\Sigma}_1&=&\frac{i\gamma M\Omega\left[(1+\eta M^2){\bf v}+M{\bf N}_{\times}{\bf v}^*\right]}{G_+\left[(1+\eta M^2)^2-M^2\left|{\bf N}_{\times}\right|^2\right]}=M{\bf B},\nonumber \\
N_1&=&-\eta M,
\eeq
where ${\bf B}$ and $\eta$ are bounded functions. 

Due to the oscillatory behaviour of the system \cite{Ringstrom3,VIII}, we introduce the following variables \cite{HHLC}
\beq
{\mbold\Sigma}_{\times}+i{\bf N}_{\times} &\equiv & e^{2i\psi}{\bf X}, \\
{\mbold\Sigma}_{\times}-i{\bf N}_{\times} &\equiv & e^{-2i\psi}{\bf Y}, \\
{\bf v} &\equiv & e^{-i\theta}v_2,
\eeq
where 
\beq
\psi' &=& \frac{\sqrt{3}}{M}, \\
\theta' &=& \frac{\sqrt{3}}{M}\Bigg\{v_1+\frac 12M\mathrm{Im}\left[(1+v_1){\bf X}e^{2i(\theta+\psi)}+ (1-v_1){\bf Y}e^{2i(\theta-\psi)}\right] \nonumber \\
&& \phantom{ \frac{\sqrt{3}}{M}\Bigg\{v_1} 
+\frac{2\gamma\eta M^3\Omega v_1}{G_+\left[(1+\eta M^2)^2-M^2\left|{\bf N}_{\times}\right|^2\right]}\Bigg\}.
\eeq
The angular variables $\psi$ and $\theta$ are introduced to take care of the rapid oscillation 
as $M\rightarrow 0$. We note that the variables $\psi$ and $\theta$ are not in 
synchronization since $v_1<1$ (asymptotically they can be if $v_1\rightarrow 1$). Hence, in general, we expect two different oscillations with different frequencies.  
 Moreover, 
we note that both of these rapid oscillations are observable; e.g., by considering the scalars: 
\beq
S_1&=&({\mbold\Sigma}_{\times}+i{\bf N}_{\times})({\mbold\Sigma}_{\times}^*+i{\bf N}_{\times}^*)=e^{4i\psi}{\bf X}{\bf Y}^*, \nonumber \\
S_2&=&({\mbold\Sigma}_{\times}^2+{\bf N}_{\times}^2)({\bf v}^*)^4=e^{4i\theta}{\bf X}{\bf Y}v_2^4. \nonumber
\eeq
We should also point out that we have replaced the two complex variables,  ${\mbold\Sigma}_{\times}$ and ${\bf N}_{\times}$ with the two complex and one real variables ${\bf X}$, ${\bf Y}$ and $\psi$; hence, care has to be used whenever we want to extract the physical quantities. 

\subsection{Reduced System} \label{sect:redsys}
From Theorem \ref{thmNbar} we have that $M\rightarrow 0$ at late times. In Appendix \ref{App:A} it is shown how the system of equations effectively reduces to a simpler set of equations at late times. As $M\rightarrow 0$ the reduced system becomes increasingly accurate and the asymptotic dynamics can be deduced from this reduced system. 
Defining 
\beq
\sigma_1&\equiv& \left|{\bf N}_{\times}\right|^2+ \left|{\mbold\Sigma}_{\times}\right|^2=\frac{1}{2}\left(|{\bf X}|^2+|{\bf Y}|^2\right),\\
\sigma_2&\equiv& 2\mathrm{Im}\left({\mbold\Sigma}_{\times}{\bf N}_{\times}^*\right)=-\frac{1}{2}\left(|{\bf X}|^2-|{\bf Y}|^2\right),
\eeq
the system of equations effectively reduces to the following system: 
\beq
\Sigma_+'&=&
(Q-2)\Sigma_++\eta-\sigma_1
+\frac{\gamma\Omega}{2G_+}\left(-2v_1^2+v_2^2\right)     \label{red:Sigma+}\\
\sigma_1'&=& 2(Q+\Sigma_+-1)\sigma_1\\ 
\eta' &=& 2(Q-\Sigma_+)\eta \\
M'&=&
-\left(Q+2\Sigma_+\right)M\\
\Omega'&=& \frac{\Omega}{G_+}\Big\{2Q-(3\gamma-2)
 +\left[2Q(\gamma-1)-(2-\gamma)-\gamma\mathcal{S}\right]V^2\Big\}
 \\
v_1' &=& \left(T+2\Sigma_+\right)v_1\\
v_2'&=&(T-\Sigma_+)v_2
\label{red:v2}
\eeq
where
\beq
Q&=& 2\Sigma^2_++\sigma_1+\frac
12\frac{(3\gamma-2)+(2-\gamma)V^2}{1+(\gamma-1)V^2}\Omega, \\
V^2\mathcal{S} &=& \left(-2v_1^2+v_2^2\right)\Sigma_+.
\eeq
These variables are subject to the constraint
\beq
1&=& \Sigma^2_++\sigma_1+\eta+\Omega.
\label{eq:redHamiltonian}\eeq
Furthermore, $\sigma_2$ is determined from
\[ 
\sigma_2=-\frac{\gamma\Omega v_1}{G_+},
\]
which gives the bound 
\beq 
\sigma_1\geq \frac{\gamma\Omega |v_1|}{G_+}.
\label{redboundv1}\eeq
Using the constraint (\ref{eq:redHamiltonian}) we can solve for $\sigma_1$ or $\Omega$. 

\subsubsection{Monotonic functions for the reduced system}
The two functions
\beq
R_1&\equiv & \frac{\beta\Omega}{\sigma_1^m(1-m\Sigma_+)^{2(1-m)}}, \quad m=\frac{1}{2}(4-3\gamma), \nonumber\\
R_1'&=& \frac{2}{(1-m\Sigma_+)}\left[(\Sigma_+-m)^2+3(1-\gamma)(1-m)\eta+\frac{m(1-m)(2-m)|{\bf v}|^2\Omega}{G_+}\right]R_1,\nonumber 
\eeq
and 
\beq
R_2&\equiv& \frac{(1-V^2)^{\frac 32(2-\gamma)}\Omega}{v_1^2G_+},  \\
R_2'&=&\left[9(1-\gamma)+(2\Sigma_+-1)^2+\sigma_1+\frac
12\frac{(3\gamma-2)+(2-\gamma)V^2}{1+(\gamma-1)V^2}\Omega\right]R_2, \nonumber
\eeq
are both monotonically increasing for $2/3<\gamma\leq 1$. 

Moreover, there are two more monotonic functions for the reduced system: 
\beq
R_3 &\equiv & \frac{v_1|{\bf v}|^2}{(1-V^2)^{\frac 32(2-\gamma)}}, \quad 
R_3' = 3(3\gamma-4)R_2,\\
R_4&\equiv &\frac{\eta\sigma_1}{\beta^2\Omega^2}, \quad R_4'=6(\gamma-1)R_4.
\eeq
To relate these functions to the full system, we note that for all the above functions we can write $R_{\alpha}=p/q$ where the functions $p$ and $q$ are both non-negative and bounded. Assume that for the reduced system, $R_{\alpha}'=F({\sf X})R_{\alpha}$. Using a similar trick as in Appendix \ref{App:A} we can show that for the full system we have 
\beq
F({\sf X})\geq 0 \quad \Rightarrow \quad F({\sf X})\rightarrow 0 \quad\text{or}\quad q\rightarrow 0. \nonumber \\
F({\sf X})\leq 0 \quad \Rightarrow \quad F({\sf X})\rightarrow 0 \quad\text{or}\quad p\rightarrow 0. \nonumber
\eeq
Hence, effectively, we can use these functions as monotonic functions for the full system at sufficently late times. 

\subsubsection{Type VII$_0$ ($\eta=0$) equilibrium points for the reduced system} There is a number of type VII$_0$ equilibrium points for the reduced system (see \cite{HHLC}). However, all of them have an unstable $\eta$-direction and can therefore not act as late-time attractors for a type VIII model.   
\subsubsection{Type VIII ($\eta>0$) equilibrium points for the reduced system}
\begin{enumerate}
\item{} $\widetilde{P}_1$: $2/3<\gamma<1$, $Q=\Sigma_+=\frac 12(3\gamma-2)$, $\eta=\frac 34(3\gamma-2)(2-\gamma)$, $\Omega=3(1-\gamma)$, $\sigma_1=V=0$. \\
Eigenvalues: 
\[ -\frac{3}{2}(2-\gamma), ~-6(1-\gamma)[\times 2], ~-\frac 34(2-\gamma)\left[1\pm\sqrt{1-\frac{8(3\gamma-2)(1-\gamma)}{2-\gamma}}\right].\]
\item{} $\widetilde{P}_2$: $2/3<\gamma<2$, $Q=\Sigma_+=\frac 12$, $\eta=\frac 34$, $\Omega=\sigma_1=V=0$. \\
Eigenvalues: 
\[ 0, ~ 3(1-\gamma),~-3(1-\gamma),~-\frac 32,~\frac 32(2\gamma-3).\]
\item{} $\widetilde{L}_1(k)$: $\gamma=1$, $Q=\Sigma_+=\frac 12$, $\eta=\frac 34$, $v_1=k$, $\Omega=\sigma_1=v_2=0$, $0<k<1$.
Eigenvalues: 
\[ 0[\times 3], ~-\frac 32[\times 2].\]
\item{} $\widetilde{L}_2(k)$: $\gamma=3/2$, $Q=\Sigma_+=\frac 12$, $\eta=\frac 34$, $v_2=k$, $\Omega=\sigma_1=v_1=0$, $0<k<1$.
Eigenvalues: 
\[ 0[\times 2], ~-\frac 32[\times 2],~\frac 32.\]
\item{} $\widetilde{E}_1$: $2/3<\gamma<2$, $Q=\Sigma_+=\frac 12$, $\eta=\frac 34$, $v_1=1$, $\Omega=\sigma_1=v_2=0$. \\
Eigenvalues: 
\[ 0[\times 2], ~-\frac 32[\times 2],~-\frac {3(\gamma-1)}{2-\gamma}.\]
\item{} $\widetilde{E}_2$: $2/3<\gamma<2$, $Q=\Sigma_+=\frac 12$, $\eta=\frac 34$, $v_2=1$, $\Omega=\sigma_1=v_1=0$. \\
Eigenvalues: 
\[ 0, ~-\frac 32[\times 2],~\frac 32,~-\frac {3(2\gamma-3)}{2(2-\gamma)}.\]
\end{enumerate}
We note that several of these equilibrium points have a number of zero eigenvalues. Further analysis is therefore required in order to determine the stability of these points. The result of this stability analysis is shown in Table \ref{tab:outline} and also in the next section where the decay rates are calculated. 

For the stability analysis and the calculation of the decay rates there are a few things worth mentioning. For the two lines of equilibria, $\widetilde{L}_1(k)$ and $\widetilde{L}_2(k)$ for $\gamma=1$ and $\gamma=3/2$ respectively, two of the zero eigenvalues of the linearised system actually correspond to a non-trivial Jordan block; i.e., one of the Jordan blocks is of the form 
\beq 
J_1=\begin{bmatrix} 0 & 1 \\ 0 & 0 \end{bmatrix}.\nonumber
\eeq
This means that a generic solution `drifts' along the line of equilibria; the amount of `drifting' depends on the second order terms. Let us, for the sake of illustration, consider the case $\gamma=1$. In this case, the equilibria $\widetilde{P}_2$, $\widetilde{L}_1(k)$, and $\widetilde{E}_1$ all have 3 zero eigenvalues, with all other eigenvalues being negative. Note also that 
\[ \lim_{k\rightarrow 0}\widetilde{L}_1(k)=\widetilde{P}_2, \quad \lim_{k\rightarrow 1}\widetilde{L}_1(k)=\widetilde{E}_1.\] 
Centre manifold theory can now be used to determine the stability and the decay rates \cite{Carr}. Since there are 3 zero eigenvalues the centre manifold is 3 dimensional. However, this centre manifold can be simplified by using the function $R_4$. Since $R_4'=0$ for $\gamma=1$ we can write 
\[ \eta\sigma_1=K\beta^2\Omega^2, \]
for a constant $K$, which can be used to reduce the centre  manifold to a more tractable 2 dimensional system. Standard techniques then reveal that $\widetilde{E}_1$ and $\widetilde{L}_1(k)$ are unstable, while $\widetilde{P}_2$ is stable (close to $\widetilde{L}_1(k)$  solutions drift towards the stable point $\widetilde{P}_2$).

\section{Late time behaviour} 
By assuming an ansatz with coefficients and exponents to be determined we can calculate the decay rate for a generic tilted type VIII model. 
For the cases where the linearised system has zero eigenvalues, an analysis of the centre manifold is needed, as explained above. For the decay rates given below, only the leading order terms are given. The decay rates are verified by numerical simulations, see Figures \ref{Fig:g=45}-\ref{Fig:g=43} below. 

The Hubble-normalised Weyl invariants $\Wm$ and $\Wc$ are defined by 
\beq
\left(\Wm,~\Wc\right)\equiv \frac{1}{48H^4}\left(C_{\alpha\beta\gamma\delta}C^{\alpha\beta\gamma\delta},~C_{\alpha\beta\gamma\delta}{}^\star C^{\alpha\beta\gamma\delta}\right),
\eeq
and are also given below in the asymptotic regimes. The variable $\psi_2$ in these expressions is defined by $S_1=|S_1|e^{2i\psi_2}$ and is related asymptotically to $\psi$ via $\psi_2\approx 2\psi+\psi_0$ where $\psi_0$ is a constant. 

\begin{figure}
\centering
\includegraphics[width=10cm]{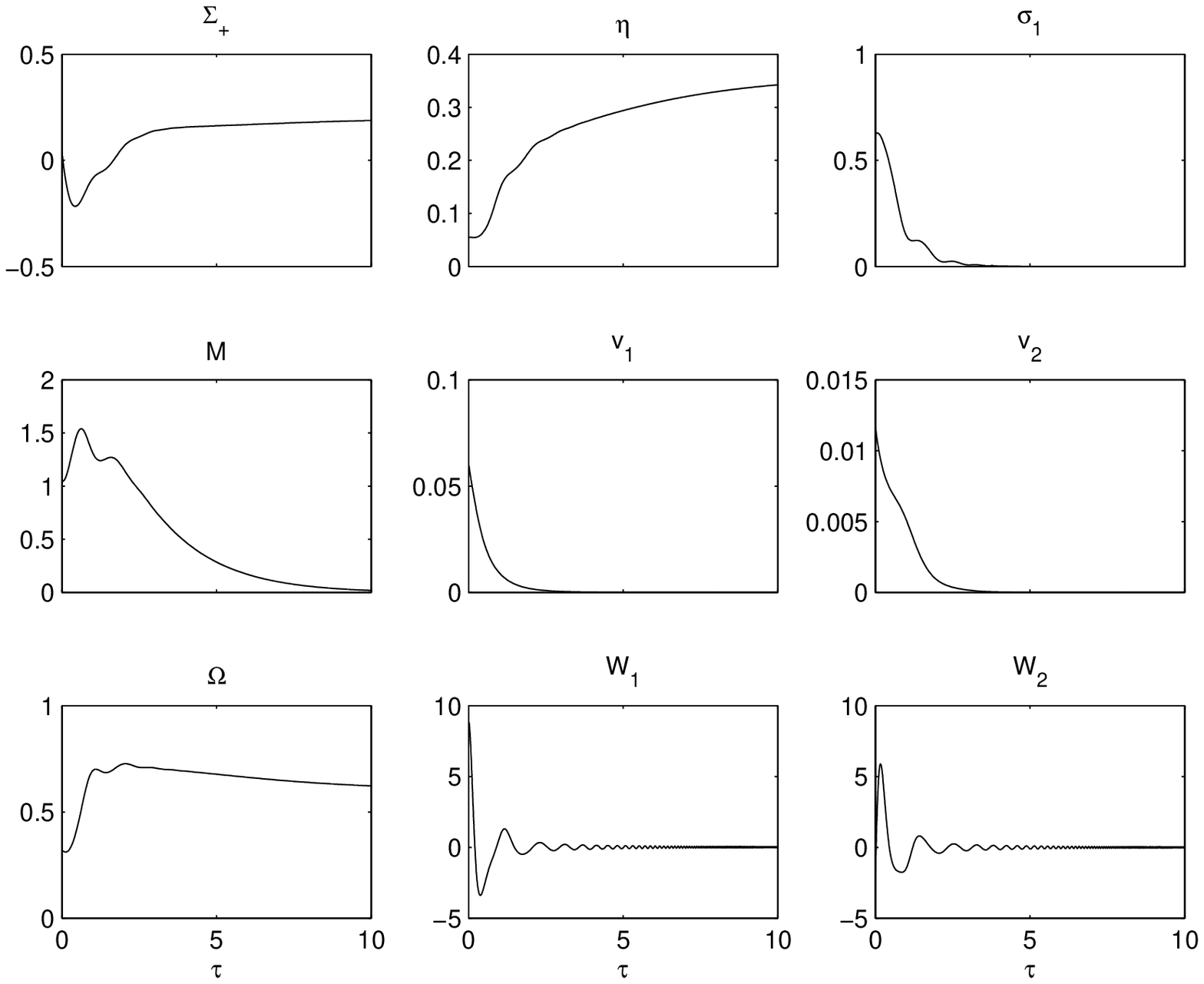}
\caption{Numerical plots of the case $\gamma=4/5$.}\label{Fig:g=45}
\end{figure}
\begin{figure}
\centering
\includegraphics[width=10cm]{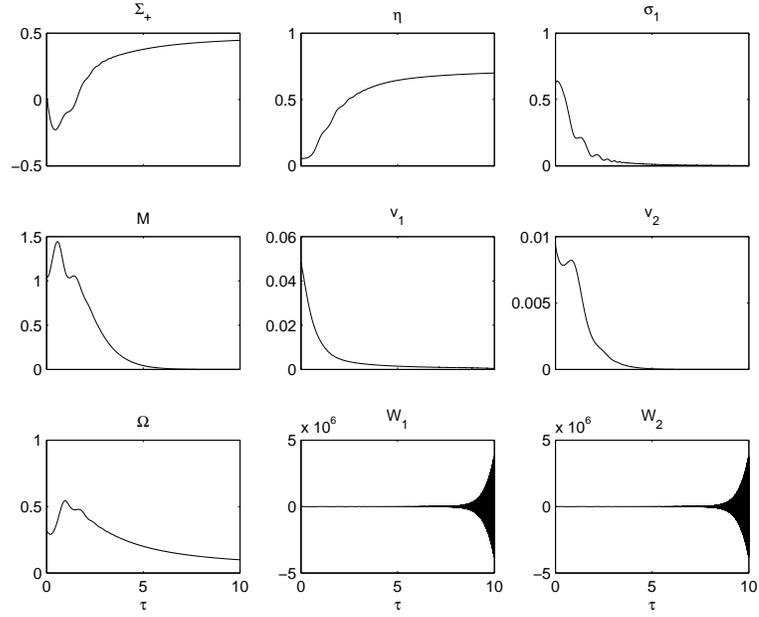}
\caption{Numerical plots of the case $\gamma=1$.}\label{Fig:g=1}
\end{figure}

\begin{figure}
\centering
\includegraphics[width=10cm]{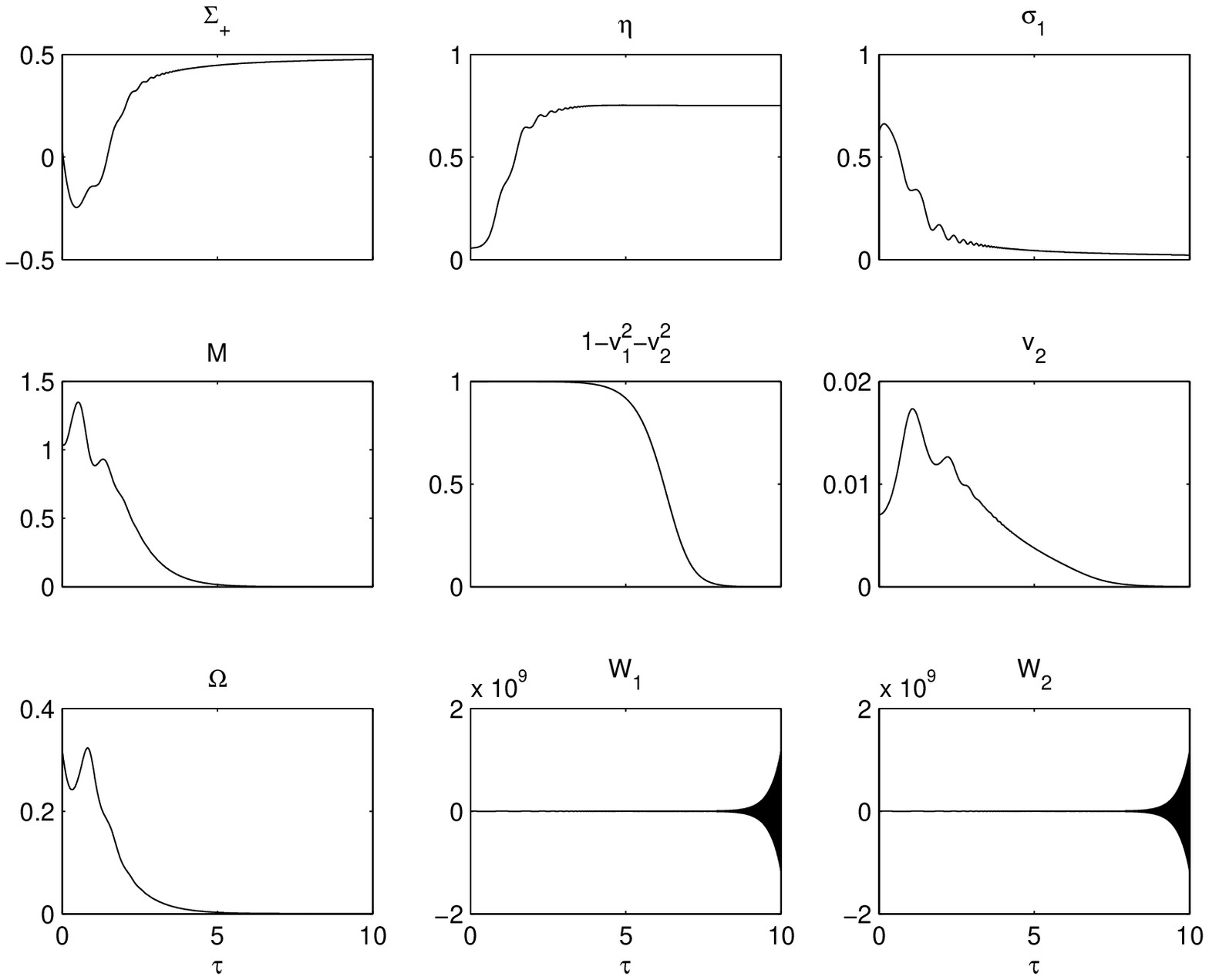}
\caption{Numerical plots of the case $\gamma=4/3$.}\label{Fig:g=43}
\end{figure}
\subsection{Decay rates for $2/3<\gamma<1$:} The attractor is $\widetilde{P}_1$ and the decay rates are: 
\beq
\Sigma_+(\tau) &\approx & \frac 12(3\gamma-2)+\mathcal{O}(e^{\lambda_i\tau}) ,\nonumber \\
\eta(\tau) &\approx & \frac 34(3\gamma-2)(2-\gamma)+\mathcal{O}(e^{\lambda_i\tau}), \nonumber \\
\sigma_1(\tau) &\approx &\hat{\sigma}_1e^{-6(1-\gamma)\tau} , \nonumber \\ 
M(\tau) &\approx &\hat{M}e^{-\frac 32(3\gamma-2)\tau} , \nonumber \\
v_1(\tau) &\approx & \hat{v}_1e^{-6(1-\gamma)\tau} ,\quad 0\leq 3\gamma(1-\gamma)\hat{v}_1\leq \hat{\sigma}_1\nonumber \\ 
v_2(\tau) &\approx &\hat{v}_2e^{-\frac 32(2-\gamma)\tau}  , \nonumber \\
\Omega(\tau)&\approx & 3(1-\gamma)+\mathcal{O}(e^{\lambda_i\tau}).
\eeq
Here, $\lambda_i$ are the eigenvalues of the linearised system. 

The variables $\psi$ and $\theta$ have the asymptotic behaviour: 
\beq
\psi&\approx & \hat\psi+\frac{2\sqrt{3}}{3\hat{M}(3\gamma-2)}e^{\frac 32(3\gamma-2)\tau}, \nonumber \\
\theta&\approx & \begin{cases} 
\hat\theta, & 2/3<\gamma<6/7, \\
\hat\theta+\frac{\sqrt{3}\hat{v}_1}{\hat{M}}\tau, & \gamma=6/7, \\
\hat\theta+\frac{2\sqrt{3}\hat{v}_1}{3\hat{M}(7\gamma-6)}e^{\frac 32(7\gamma-6)\tau}, & 6/7<\gamma<1,
\end{cases}
\eeq
Asymptotic behaviour of the Weyl tensor, $\hat{\sigma}_1\neq 3\gamma(1-\gamma)\hat{v}_1$: 
\beq
(\Wm,\Wc)\approx \begin{cases} (0,0), \quad & 2/3<\gamma<4/5, \\
	-\frac{12}{\hat{M}^2}\sqrt{\hat{\sigma}_1^2-9\gamma^2(1-\gamma)^2\hat{v}_1^2}\ (\cos2\psi_2,\sin2\psi_2), \quad & \gamma=4/5, \\
	-\frac{12}{\hat{M}^2}\sqrt{\hat{\sigma}_1^2-9\gamma^2(1-\gamma)^2\hat{v}_1^2}\ e^{3(5\gamma-4)\tau}(\cos2\psi_2,\sin2\psi_2), \quad & 4/5<\gamma<1.
\end{cases}\nonumber \\
\eeq

\subsection{Decay rates for $\gamma=1$:} The attractor is $\widetilde{P}_2$ and the decay rates are: 
\beq
\Sigma_+(\tau) &\approx &\frac 12\left[1-\frac 1\tau \right], \nonumber \\
\eta(\tau) &\approx &\frac 34\left[1-\frac{2}{3\tau}\right], \nonumber \\
\sigma_1(\tau) &\approx &\frac{\hat{\sigma}_1}{\tau^2} , \nonumber \\ 
M(\tau) &\approx &\hat{M}\tau^{\frac 32}e^{-\frac 32\tau} , \nonumber \\
v_1(\tau) &\approx &\frac{\hat{v}_1}{\tau} , \quad |\hat{v}_1|\leq \hat{\sigma}_1,\nonumber \\ 
v_2(\tau) &\approx &\hat{v}_2\tau^{\frac 12}e^{-\frac 32\tau}  , \nonumber \\
\Omega(\tau)&\approx &\frac 1\tau.
\eeq
The variables $\psi$ and $\theta$ have the asymptotic behaviour: 
\beq
\psi&\approx & \hat\psi+\frac{2\sqrt{3}}{3\hat{M}}\tau^{-\frac 32}e^{\frac 32\tau}, \nonumber \\
\theta&\approx & 
\hat\theta+\frac{2\sqrt{3}\hat{v}_1}{3\hat{M}}\tau^{-\frac 52}e^{\frac 32\tau}.
\eeq
Asymptotic behaviour of the Weyl tensor, $\hat{\sigma}_1\neq \hat{v}_1$: 
\beq
(\Wm,\Wc)\approx -\frac{12}{\hat{M}^2}\sqrt{\hat{\sigma}_1^2-\hat{v}_1^2}\ \tau^{-5}e^{3\tau}(\cos2\psi_2,\sin2\psi_2).
\eeq

\subsection{Decay rates for $1<\gamma<2$:} The attractor is $\widetilde{E}_1$ and the decay rates are: 
\beq
\Sigma_+(\tau) &\approx &\frac 12\left[1-\frac 1{2\tau} \right], \nonumber \\
\eta(\tau) &\approx &\frac 34\left[1+\frac{1}{12\tau^2}\right], \nonumber \\
\sigma_1(\tau) &\approx &\frac{\hat{\sigma}_1}{\tau}, \quad \frac 18\leq \hat{\sigma}_1<\frac 14, \nonumber \\ 
M(\tau) &\approx &\hat{M}\tau^{\frac 34}e^{-\frac 32\tau} , \nonumber \\
v_1^2(\tau) &\approx &1-\hat{C}^2\tau^{\frac{1}{2-\gamma}}e^{-\frac{6(\gamma-1)}{2-\gamma}\tau}-v_2^2(\tau) , \nonumber \\ 
v_2(\tau) &\approx &\hat{v}_2\tau^{\frac 34}e^{-\frac 32\tau}  , \nonumber \\
\Omega(\tau)&\approx &\left(\frac 14-\hat{\sigma}_1\right)\frac 1\tau.
\eeq
The variables $\psi$ and $\theta$ have the asymptotic behaviour: 
\beq
\psi&\approx & \hat\psi+\frac{2\sqrt{3}}{3\hat{M}}\tau^{-\frac 34}e^{\frac 32\tau}, \nonumber \\
\theta&\approx & 
\hat\theta+\frac{2\sqrt{3}}{3\hat{M}}\tau^{-\frac 34}e^{\frac 32\tau}.
\eeq
Asymptotic behaviour of the Weyl tensor, $\hat{\sigma}_1\neq 1/8$: 
\beq
(\Wm,\Wc)\approx -\frac{3}{\hat{M}^2}\sqrt{8\hat{\sigma}_1-1}\ \tau^{-\frac 52}e^{3\tau}(\cos2\psi_2,\sin2\psi_2).
\eeq

\section{Discussion} 
\begin{table}
\centering
\begin{tabular}{|c|c|c|l|}
\hline
 Invariant &   &  &  \\
 subspace & Matter & Attractor & Asymptotic tilt \\ \hline \hline
 $T(VIII)$ & $2/3<\gamma<1$ & $\widetilde{P}_1$ & non-tilted (e)\\
           & $\gamma=1$ & $\widetilde{P}_2$ & non-tilted (p) \\
	    & $1<\gamma<2$ & $\widetilde{E}_1$ & extremely tilted (e) \\
	    \hline
$N(VIII)$ & $2/3<\gamma<1$ & $\widetilde{P}_1$ & non-tilted (e)\\
           & $\gamma=1$ & $\widetilde{P}_2$ & non-tilted (p) \\
	    & $1<\gamma<2$ & $\widetilde{E}_1$ & extremely tilted (e) \\
	    \hline
$F(VIII)$ & $2/3<\gamma<1$ & $\widetilde{P}_1$ & non-tilted (e) \\
           & $1\leq\gamma<3/2$ & $\widetilde{P}_2$ & non-tilted (e) \\
	   & $\gamma=3/2$ & $\widetilde{E}_2$ & extremely tilted (p) \\
	    & $3/2<\gamma<2$ & $\widetilde{E}_2$ & extremely tilted (e) \\
	    \hline
$T_1(VIII)$ &$2/3<\gamma<1$ & $\widetilde{P}_1$ & non-tilted (e)  \\
           & $\gamma=1$ & $\widetilde{P}_2$ & non-tilted (p) \\
	    & $1<\gamma<2$ & $\widetilde{E}_1$ & extremely tilted (e) \\
\hline
$B(VIII)$ &  $2/3<\gamma<1$ & $\widetilde{P}_1$ & no tilt\\
           & $1\leq\gamma<2$ & $\widetilde{P}_2$ & no tilt\\
\hline
\end{tabular}
\caption{The late-time behaviour of the VIII Bianchi  model with a tilted
$\gamma$-law perfect fluid (see the text for details and
references). The comments refer to the late-time asymptotics, and for all cases 
$\bar{N}\rightarrow \infty$. The case $0<\gamma<2/3$ is covered by the no-hair theorem (the non-tilted version is given in \cite{Wald}, the tilted version in \cite{CH}). The right-most column indicates the asymptotic tilt and whether the tilt velocity, $V$, approaches this state exponentially (e) or power law (p) in terms of the dynamical time $\tau$.} \label{tab:outline}
\end{table}

Table \ref{tab:outline} displays an outline of the late-time asymptotic behaviour of tilted Bianchi type VIII models. As for the late-time behaviour of the Bianchi models, all of the non-type-IX class A models have now been studied. The only remaining class A model, namely the closed type IX model is notoriously difficult and requires a different formalism. Of the class B models, an obvious lacuna is the type VI$_h$ model, all other tilted class B models have been studied. 

Regarding the tilted type VIII models we have seen that the extreme Weyl-curvature dominance (in the terminology of \cite{BHWeyl}) for $4/5<\gamma<2$ which was found for the non-tilted models, persists into the tilted model. This extreme Weyl-curvature dominance is a signal of the self-similarity breaking that occurs at late times for these models. More explicitly, this Weyl-curvature dominance is a result of an increasingly rapid oscillation that takes place in the shear and the curvature variables. This rapid oscillation is exactly what prevents the type VIII models to be asymptotically self-similar. 

Moreover, we have shown that for the type VIII models with fluids stiffer than dust ($1<\gamma<2$) the tilt becomes asymptotically extreme at late times. The energy density itself asymptotically approaches vacuum but does so rather slowly, 
\[ \Omega\propto\tau^{-1}\sim (\ln t)^{-1}.\] 
We can compare this slow decay with the non-tilted case for which: 
\[ \text{non-tilted VIII:} \quad \Omega\propto \tau^{-\frac 12}e^{-3(\gamma-1)\tau}\sim {t^{-2(\gamma-1)}(\ln t)^{-\frac\gamma 2}}.\]  
In particular, this means that for a radiation dominated tilted Bianchi type VIII model the expansion-normalised energy density, $\Omega$, only decays logarithmically at late times, in terms of the cosmological time, $t$. 

In this paper we have discussed the late-time behaviour of Bianchi type VIII cosmologies. Regarding the early-time behaviour, the models undergo Mixmaster dynamics \cite{Ringstrom1,Jantzen}, which is also the typical behaviour for inhomogeneous models \cite{UvEWE,Garfinkle}. While the early-time behaviour is similar in inhomogeneous generalisation of Bianchi models, we do not expect this for the late-time behaviour due to the dominance of inhomogeneity at late times (see, e.g., \cite{DS1,HC}).

\section*{Acknowledgments}
We would like to thank L. Hervik for useful comments on the manuscript. 
SH was supported by a Killam Postdoctoral Fellowship. 
\appendix

\section{New variables}
\label{App:A}
We  introduce the following variables
\beq
{\mbold\Sigma}_{\times}+i{\bf N}_{\times} &\equiv & e^{2i\psi}{\bf X}, \\
{\mbold\Sigma}_{\times}-i{\bf N}_{\times} &\equiv & e^{-2i\psi}{\bf Y}, \\
{\bf v} &\equiv & e^{-i\theta}v_2,
\eeq
where 
\beq
\psi' &=& \frac{\sqrt{3}}{M},\label{Newpsi}\\
\theta' &=&
\frac{\sqrt{3}}{M}\Bigg\{v_1+\frac 12M\mathrm{Im}\left[(1+v_1){\bf X}e^{2i(\theta+\psi)}+ (1-v_1){\bf Y}e^{2i(\theta-\psi)}\right] \nonumber \\
&& \phantom{ \frac{\sqrt{3}}{M}\Bigg\{v_1} 
+\frac{2\gamma\eta M^3\Omega v_1}{G_+\left[(1+\eta M^2)^2-M^2\left|{\bf N}_{\times}\right|^2\right]}\Bigg\}.
\eeq
From the remaining freedom we have in choosing the variables and the gauge function, we can choose the \emph{initial} values for ${\bf X}$ and ${\bf Y}$ to both be real. 
In any case, objects like $|{\bf X}|^2$ and $|{\bf Y}|^2$ are gauge-independent and are consequently
independent of any such choice.  We also define 
\beq M\equiv \frac 1{\bar{N}}, \quad \eta=-N_1\bar{N}.\eeq
The equations of motion are now
\beq
\Sigma_+'&=&
(Q-2)\Sigma_++\eta-\frac 12\left(|{\bf X}|^2+|{\bf Y}|^2\right)
+\frac{\gamma\Omega}{2G_+}\left(-2v_1^2+v_2^2\right) \nonumber \\
&& +{3}\mathrm{Re}\left({\bf R}^*{\mbold\Sigma}_1\right)+(\Sigma_++1)\mathrm{Re}\left({\bf X}^*{\bf Y}e^{-4i\psi}\right),\label{NewSigma+} \\
{\bf X}' &=& (Q+\Sigma_+-1){\bf X}+\sqrt{3}{\mbold\Sigma}_1{\bf R}e^{-2i\psi}+\frac{\sqrt{3}}{2}iM\eta\left({\bf X}-{\bf Y}e^{-4i\psi}\right)\nonumber \\
&& +\mathrm{Re}\left({\bf X}^*{\bf Y}e^{-4i\psi}\right){\bf X}-(1+\Sigma_+){\bf Y}e^{-4i\psi}+\frac{\sqrt{3}\gamma\Omega v_2^2}{2G_+}e^{-2i(\theta+\psi)},\\
{\bf Y}' &=& (Q+\Sigma_+-1){\bf Y}+\sqrt{3}{\mbold\Sigma}_1{\bf R}e^{2i\psi}+\frac{\sqrt{3}}{2}iM\eta\left({\bf X}e^{4i\psi}-{\bf Y}\right)\nonumber \\
&& +\mathrm{Re}\left({\bf X}^*{\bf Y}e^{-4i\psi}\right){\bf Y}-(1+\Sigma_+){\bf X}e^{4i\psi}+\frac{\sqrt{3}\gamma\Omega v_2^2}{2G_+}e^{-2i(\theta-\psi)}, \\
M'&=&
-M\left[Q+2\Sigma_+
+\mathrm{Re}\left({\bf X}^*{\bf Y}e^{-4i\psi}\right)+\sqrt{3}M\mathrm{Im}\left({\bf Y}^*{\bf
X}e^{4i\psi}\right)\right],\\
\eta'&=&
\eta\left[2Q-2\Sigma_+
+2\mathrm{Re}\left({\bf X}^*{\bf Y}e^{-4i\psi}\right)+\sqrt{3}M\mathrm{Im}\left({\bf Y}^*{\bf
X}e^{4i\psi}\right)\right],\\
\Omega'&=& \frac{\Omega}{G_+}\Big\{2Q-(3\gamma-2)
 +\left[2Q(\gamma-1)-(2-\gamma)-\gamma\mathcal{S}\right]V^2\Big\} \nonumber \\
 && +2\mathrm{Re}\left({\bf X}^*{\bf Y}e^{-4i\psi}\right)\Omega \label{NewOmega}
 \\
v_1' &=& \left(T+2\Sigma_+\right)v_1-\sqrt{3}\mathrm{Re}\left[({\mbold\Sigma}_{1}+{\bf R})e^{i\theta}\right]v_2 \nonumber \\
&& +\frac{\sqrt{3}}{2}\mathrm{Re}\left({\bf X}e^{2i(\theta+\psi)}-{\bf Y}e^{2i(\theta-\psi)}\right)v_2^2\\
v_2'&=&\left\{T-\Sigma_+-\frac{\sqrt{3}}{2}\mathrm{Re}\left[(1+v_1){\bf X}e^{2i(\theta+\psi)}+(1-v_1){\bf Y}e^{2i(\theta-\psi)}\right]\right\}v_2,\nonumber  \\ 
&& -\sqrt{3}(\mbold{\Sigma}_1-{\bf R})e^{i\theta}v_1\label{Newv2}
\eeq
where
\beq
Q&=& 2(\Sigma^2_++|{\mbold\Sigma}_1|^2)+\frac{1}{2}\left(|{\bf X}|^2+|{\bf Y}|^2\right)+\frac
12\frac{(3\gamma-2)+(2-\gamma)V^2}{1+(\gamma-1)V^2}\Omega.
\eeq
These variables are subject to the constraints
\beq
1&=& \Sigma^2_++\left|{\mbold\Sigma}_1\right|^2+\frac{1}{4}M^2\eta^2+\eta+\frac{1}{2}\left(|{\bf X}|^2+|{\bf Y}|^2\right)+\Omega \\
0 &=& \frac{1}{2}\left(|{\bf X}|^2-|{\bf Y}|^2\right)-\frac{\gamma\Omega v_1}{G_+}. \label{NewConst3}
\eeq 
We can also split $V^2{\mathcal{S}}$ and $T$ into oscillatory, and non-oscillatory parts: 
\beq
V^2\mathcal{S}&=& \Sigma_+\left(-2v_1^2+v_2^2\right)\nonumber \\ && 
+\frac{\sqrt{3}}{2}\mathrm{Re}\left[\left({\bf X}e^{2i(\theta+\psi)}+{\bf Y}e^{2i(\theta-\psi)}\right)v_2^2+4{\mbold\Sigma_1}e^{i\theta}v_1v_2\right],\\
T & =& \frac{(3\gamma-4)(1-V^2)+(2-\gamma)\Sigma_+\left(-2v_1^2+v_2^2\right)}{1-(\gamma-1)V^2}\nonumber\\
&&+ \frac{\sqrt{3}(2-\gamma)}{2(1-(\gamma-1)V^2)}\mathrm{Re}\left[\left({\bf X}e^{2i(\theta+\psi)}+{\bf Y}e^{2i(\theta-\psi)}\right)v_2^2+4{\mbold\Sigma_1}e^{i\theta}v_1v_2\right].
\eeq
\subsection{From the full to the reduced system}
From Theorem \ref{thmNbar} we have that $M\rightarrow 0$. This implies that there exists an $\epsilon(\tau)$ such that $M(\tau)=\epsilon^{p+1}(\tau)$ where $p$ is some positive constant. Moreover, define $\delta(\tau)\equiv 1-v_1^2$.

At a time $\tau$ where $\epsilon$ is sufficiently small, we split into two cases. 
\subsubsection{$\delta\geq \epsilon^p$:} We perform a change of variables given by a set of functions $h^i(x^j)$:
\[  \hat{x}^i=h^i(x^j)=\frac{x^i}{1+Mf}-Mg, \quad f,g \text{ bounded.}\]
 We choose the functions such that 
\[ {\hat{x}^i}{}'  =F^i(x^j)+MB^i(x^j)\]
where $B^i(x^j)$ are bounded and $F^i(x^j)$ contain no terms with $e^{-4i\psi}$, $e^{i(\theta+\psi)}$ etc.. We note that the jacobian ${\bf J}\equiv(\partial h^i/\partial x^j)$ has determinant: 
\beq
\det({\bf J})=1+\sum_{k=1}^nb_k\left(\frac{M}{1-v_1^2}\right)^k, \quad b_k \text{ bounded.}
\eeq
Furthermore,
\[ \frac{M}{1-v_1^2}=\frac{\epsilon^{p+1}}{\delta}\leq \epsilon;\]
therefore, for sufficiently small  $\epsilon$ the inverse function theorem implies that the maps $h^i(x^j)$ have a continuous and well-defined inverse. This further means that 
\beq
{\hat{x}^i}{}'  =\hat{F}^i(\hat{x}^j)+\hat{M}\hat{B}^i(x^j)+\mathcal{O}(\epsilon).
\eeq
\subsubsection{ $\delta<\epsilon^p$:} We now choose functions $h^i(x^j)$ such that 
\[ {\hat{x}^i}{}'  =F^i(x^j)+MB^i(x^j)+v_2^2\tilde{B}^i(x_j),\]
where $B^i(x^j)$, $\tilde{B}^i(x^j)$ are bounded and $F^i(x^j)$ contain no terms with $e^{-4i\psi}$, $e^{i(\theta+\psi)}$ etc.. In this case we get 
\beq
\det({\bf J})=1+\sum_{k=1}^n\tilde{b}_k{M}^k, \quad \tilde{b}_k \text{ bounded.}
\eeq
So for sufficiently small $\epsilon$ the maps $h^i(x^j)$ have a continuous and well-defined inverse. Moreover, $v_2^2=V^2-v_1^2\leq \delta\leq \epsilon^p$. Hence, 
\beq
{\hat{x}^i}{}'  =\hat{F}^i(\hat{x}^j)+\hat{M}\hat{B}^i(x^j)+\mathcal{O}(\epsilon^p).
\eeq
Both of these cases reduce to the same reduced system. 

We should point out that in the asymptotic regime we can obtain tighter bounds on the oscillatory terms using, for example, the linearised analysis in the tilted type VII$_0$ paper \cite{HHLC}. 

\newpage
\bibliographystyle{amsplain}

\end{document}